\documentclass{PoS}

\title{Medium Effects in Parton Distributions\footnote{NT@UW-11-32}}

\ShortTitle{Medium Effects in Parton Distributions}

\author{\speaker{William Detmold}\\
  Department of Physics, College of William \& Mary,
  Williamsburg, VA 23187, USA\\
  Jefferson Lab, 12000 Jefferson Ave, Newport News, VA 23606, USA \\
 E-mail:
  \email{wdetmold@jlab.org}}

\author{Huey-wen Lin\\
  Department of Physics, University of Washington, Seattle, WA 98195, USA\\
  E-mail: \email{hwlin@phys.washington.edu}}

\abstract{Understanding the effects of a background hadronic medium on
  hadronic observables is important in the context of hadron
  structure. Many experiments probing nucleon structure make use of
  nuclear targets and unraveling the modifications that ensue is a
  complex task. Using lattice QCD, we investigate the ab initio
  computation of hadron structure in a medium, focusing on the
  structure of the pion in a Bose-condensed gas of pions.  }

\FullConference{ The XXIX International Symposium on Lattice Field Theory - Lattice 2011\\
  July 10-16, 2011\\
  Squaw Valley, Lake Tahoe, California}

\begin{document}

\section{Introduction}

An important part of our experimental understanding of hadron
substructure is that hadrons in a medium are different to those
in free space. The ``EMC effect'', the modification of the proton
$F_2(x,Q^2)$ structure function inside a nucleus (as first observed in
1983 by the European Muon Collaboration \cite{Aubert:1983xm}) is one
of the most famous examples of medium sensitivity. This effect has
been studied in many hadronic models with varying degrees of success
(see Ref.~\cite{Norton:2003cb} for a recent review) but an
understanding from first principles is lacking. While the EMC effect
and other such modifications of hadron properties are not {\it a
  posteriori} unexpected from the viewpoint of Quantum Chromodynamics
(QCD), the complexity of QCD calculations involving nuclei has
prevented the direct investigation of such effects.  In QCD, we expect
that medium modification is a ubiquitous feature of complex hadronic
systems and, in these proceedings, we report on the study of a close
analogue of the EMC effect, namely the modification of pion structure
in the presence of a Bose condensed medium of pions. 

Since hadronic structure is a low energy consequence
of QCD, the only tool with which to perform {\it ab initio} studies
is lattice QCD (LQCD).  Lattice QCD is formulated in Euclidean
space, so physics defined on the light-cone, such as that embodied in the parton
distributions and structure functions of deep inelastic scattering, is
very difficult to address directly (an alternate approach is suggested
in Ref.~\cite{Detmold:2005gg}).  However, through the Wilsonian
operator product expansion (OPE), the Mellin moments, $\langle
x^n\rangle_h(\mu)=\int_{-1}^{1}dx\, x^n q_h(x;\mu)$, of the unpolarised
quark distribution, $q_h(x;\mu)$, in a hadron $h$ correspond to the
forward matrix elements of local operators ($\mu$ is the
renormalisation scale).  Here, we focus on the
leading twist, unpolarised operators and have
\begin{eqnarray}
  \label{eq:1}
  \langle h;\ p | {\cal O}_{q}^{\{\mu_0\ldots\mu_n\}} | h;\ p \rangle = 
\langle x^n\rangle_h p^{\mu_0}\ldots p^{\mu_n}\,,\\
{\cal O}_q^{\{\mu_0\ldots\mu_n\}}(x) = \overline{q}(x) \gamma^{\{\mu_0}
D^{\mu_1}\ldots D^{\mu_n\}} q(x)\,.
\end{eqnarray}
where $\{\ldots\}$ indicates symmetrisation of enclosed indices and
subtraction of traces. The dependence of the
various quantities on the renormalisation scale is suppressed for 
concision and $q$ represents a
particular flavour of quark field. The hadron $h$ can be a proton or
pion or it can be a more complex object such as a nucleus or a
collection of mesons. In these proceedings, matrix elements of the
$n=1$ operator are investigated in systems of up to twelve
pions.

\section{Lattice Methods}

These matrix elements can be computed using the lattice approach by
measuring two- and three- point correlation functions in the
appropriate hadronic states. For clarity, we will consider the case of
the up quark distribution of the $\pi^+$ ($u\overline{d}$) in a medium
of $\pi^+$'s. Two point functions from a source location $x_0=({\bf
  x}_0,t_0)$,
\begin{eqnarray}
  C_{m}(t,{\bf p})&=&\Big\langle 0\Big| \left[ \prod_{i=1}^{m}\sum_{\bf x}
    e^{i{\bf p}_i\cdot{\bf x}} \pi^+({\bf
      x},t)\right] \left[\pi^-(x_0)\right]^m\Big|0\Big\rangle,
       \label{eq:3}
\end{eqnarray}
where $\pi^+=\overline u \gamma_5 d$, allow the energies of systems of
$m$-pions to be determined from the dependence on Euclidean time.  The
total momentum of the $m$-pion state is ${\bf p}=\sum_{i=1}^{m}{\bf
  p}_i$ as selected by the summations over spatial sink locations (the
individual ${\bf p}_i$ are not quantum numbers). In the current
context, we shall only consider ${\bf p}={\bf 0}$.

For $t_0=0$, the spectral decomposition of the correlator has the form
\cite{Detmold:2011kw}
\begin{eqnarray}
  \label{eq:33}
C_{m}(t,{\bf 0})\to
\sum_{\ell=0}^m{\footnotesize\left(\begin{array}{c}m
      \\\ell\end{array}\right)} 
Z_m^{(\ell)} e^{-E_{m-\ell} t}e^{-E_\ell(T-t)}
   + \ldots, 
\end{eqnarray}
where $T$ is the temporal extent of the lattice and the ellipsis
denotes excited states that are exponentially suppressed as $t$
increases.  The factors $Z_m^{(\ell)}\sim|\langle \ell | (\pi^+)^m|
m-\ell\rangle|^2$ represent the overlap of the $m$-pion interpolating
operator onto $(m-\ell)$ $\pi^+$'s going forward in time and $\ell$
$\pi^-$'s going backward in time (in the ground state, the momentum of
the forward and backward going collections of pions separately
vanish). Note that $Z_m^{(m-\ell)}=Z_m^{(\ell)}$. In the limit of a
large temporal extent of the lattice ($T\to\infty$), only the term
with $\ell=0$ contributes, but at finite $T$ (corresponding to
non-zero temperature), thermal states in which some number of pions
travels around the temporal boundary are important as we shall see
below.

Corresponding three point correlation functions allow the matrix
elements in Eq.~(\ref{eq:1}) to be determined for the operator ${\cal
  O}^{(n)}_q\equiv{\cal O}_q^{\{\mu_0\ldots\mu_n\}}$,
\begin{eqnarray}
  \label{eq:4}
  C^{(n)}_{m}(\tau,t,{\bf p}) =
\Big\langle 0\Big|  \left[ \prod_{i=1}^{m}\sum_{\bf x}
    e^{i{\bf p}_i\cdot{\bf x}} \pi^+({\bf
      x},t)\right]  
 \sum_{\bf y} {\cal O}^{\{\mu_0\ldots\mu_n\}}_{u}({\bf
  y},\tau)  \left[\pi^-(x_0)\right]^m\Big|0\Big\rangle\,.
\end{eqnarray}
Here, the operator is inserted at time-slice $\tau$ injecting zero
momentum.  The spectral decomposition of the three point correlator
for ${\bf p}={\bf 0}$ is
\begin{eqnarray}
C^{(n)}_{m}(\tau,t,{\bf 0}) &=& \sum_{\ell=0}^m{\footnotesize
  \left(\begin{array}{c}m \\\ell\end{array}\right)} 
Z_m^{(\ell)} \langle {\cal O}^{(n)}_{m-\ell}\rangle
e^{-E_{m-\ell} t}e^{-E_\ell(T-t)}   + \ldots,
\label{eq:44}
\end{eqnarray}
and $ \langle {\cal O}^{(n)}_{m}\rangle= \langle m|{\cal
  O}^{(n)}_u|m\rangle$ is the matrix element of the operator in the
$m$-pion state. Excited states are suppressed in this expression and
we assume $\left\langle {\cal O}^{(n)}_{0}\right\rangle=0$ for the
cases we consider. For $m$-pion systems, contributions involving
colour singlet sub-states propagating around the temporal boundary
result in there being contributions from matrix elements of states
with $\ell<m$ pions. Near the middle of the temporal extent, $t\sim
T/2$, these contributions may be important.

In the large $T$ limit, only the $\ell=0$ state contributes in
Eqs.~(\ref{eq:33}) and (\ref{eq:44}) and it is clear that the ratio
\begin{eqnarray}
  \label{eq:5}
 R^{(n)}_m= \frac{C^{(n)}_m(\tau,t,{\bf 0})}{C_m(t,{\bf 0})}&\longrightarrow&
\langle m|{\cal O}^{\{\mu_0\ldots\mu_n\}}_{u}(\mu) | m \rangle \nonumber \\
&=&p^{\mu_0}\ldots p^{\mu_n}\langle x^n\rangle_{m\pi}(\mu)\,,
\end{eqnarray}
where $p^\mu=({\bf 0},E_m)$, This ratio will be independent of the
sink and operator insertion times, $t$ and $\tau$, provided $t_0 \ll
\tau,\ |t-\tau| \ll T$ and determines the in-medium Mellin moment.  It
follows that the double ratio ${\cal R}_m^{(n)}=R^{(n)}_m/R^{(n)}_1$
determines the ratio of moments in medium to those in free space that
we are interested in up to a simple kinematic factor. 
This double ratio
is independent of the renormalisation scale, $\mu$, obviating the need
for calculating the coefficients necessary to match lattice operators
to operators in the $\overline{\rm MS}$ scheme. 
If $t-t_0$ is not much less than $T$, the above ratios will be
contaminated by thermal contributions. However, Eqs. (\ref{eq:44}) and
(\ref{eq:33}) can still be used to extract the Mellin moments $\langle
x^n\rangle_{m\pi}(\mu)$.

In terms of a lattice calculation, the two and three point functions we
require are complicated by the many Wick contractions that arise in
multi-pion systems.  We proceed by defining partly contracted objects
\begin{eqnarray}
  \label{eq:7}
  \Pi(t;{\bf p}) &=& \sum_{\bf x} 
  S_u({\bf x},t; {\bf x}_0,t_0) S_d^\dagger({\bf x},t; {\bf
      x}_0,t_0)\,, \nonumber \\
  \widehat\Pi^{(n)}(t,\tau;{\bf p}) &=& \sum_{{\bf x},{\bf y}} e^{i{\bf p}\cdot{\bf x}} 
  S_u({\bf y},\tau;{\bf x}_0,t_0)\gamma_5\gamma^{\{\mu_0} 
 D^{\mu_1}\ldots D^{\mu_n\}}\gamma_5
  S_u({\bf x},t; {\bf y},\tau) S_d^\dagger({\bf x},t; {\bf
      x}_0,t_0)\,, \nonumber 
\end{eqnarray}
where $S_q$ is the quark propagator of flavour $q$ and we have used
$\gamma_5$-hermiticity, $S(x,y)=\gamma_5 S^\dagger(y,x)\gamma_5$, to
reverse the arguments of propagators from sink to source and
contraction on spin and colour indices is assumed. These objects are
can be viewed as time-dependent $12\times12$ matrices in spin and
colour space.  Using the techniques developed in
Refs.~\cite{Detmold:2011kw,Beane:2007es,Detmold:2008fn,Detmold:2010au},
the required contractions can now be built in terms of spin-colour
traces of products of these matrices that can be efficiently computed.
 As the
multi-hadron correlators decay very rapidly with Euclidean time, high
precision arithmetic is needed in these calculations for which we use 
the {\tt QD} library \cite{QD}.

\emph{Numerical details}: Our calculations are performed using gauge
configurations generated by the MILC collaboration
\cite{Bernard:2001av} using the rooted-staggered formulation of quarks
and the asqtad gauge action. One level of HYP smearing
\cite{Hasenfratz:2001hp} is applied to these configurations to reduce
short distance fluctuations.  The ensembles used in this study are
shown in Table \ref{tab:t1}, where we also report the number of
configurations used and the number of source locations used on each
configuration (the L ensemble has been used solely for checks of
volume dependence). Domain-wall \cite{Kaplan:1992bt,Furman:1994ky}
valence quark propagators have been calculated from APE smeared
\cite{Albanese:1987ds} sources at various locations by the NPLQCD and
LHP collaborations \cite{Beane:2011zm,WalkerLoud:2008bp}. These are
then APE smeared on the sink time-slice with fixed momentum and used
as the source for the sequential propagators connecting to the
operator generated using the same action. The source--sink separation
in the three point correlation functions is chosen at $t_{\rm
  sep}/a\equiv(t-t_0)/a \in \{16,\,20,\,24,\, 28,\,32\}$ on the C1, C2 and C3
ensembles and at $t_{\rm sep}/a\in\{24,\,32,\,48\}$ on the F ensemble. The
operator insertion time is varied over the entire lattice.

\begin{table}
  \centering
  \begin{tabular}{cccccccc}
 Label &  $a$ [fm] 	&  $L^3\times T$    & $m_\pi$ [MeV]  & $m_\pi
 L$ &  $m_\pi T$ &  Measurements \\
\hline
F & 0.09   & $28^3\times96$  & 320  & 4.1 & 14.0 & 432  \\ 
C1 & 0.12 & $20^3\times64$  & 290   & 3.7 & 11.7 &1450  \\
C2 & 0.12 & $20^3\times64$  & 350   & 4.4 & 14.2 & 837  \\
C3 & 0.12 & $20^3\times64$  & 490   & 6.2 & 19.9 & 1000  \\
L & 0.12 & $20^3\times64$  & 350   & 6.3  & 14.2 & 250
  \end{tabular}
  \caption{Details of measurements and ensembles used in this
    calculation. For each data set, the remaining columns correspond to the
    lattice spacing, lattice dimensions, valence pion mass and number of
    configurations $\times$ number of sources on each configuration. }
  \label{tab:t1}
\end{table}

\section{Thermal contamination}

\begin{figure}[!th]
  \centering
  \includegraphics[width=0.68\columnwidth]{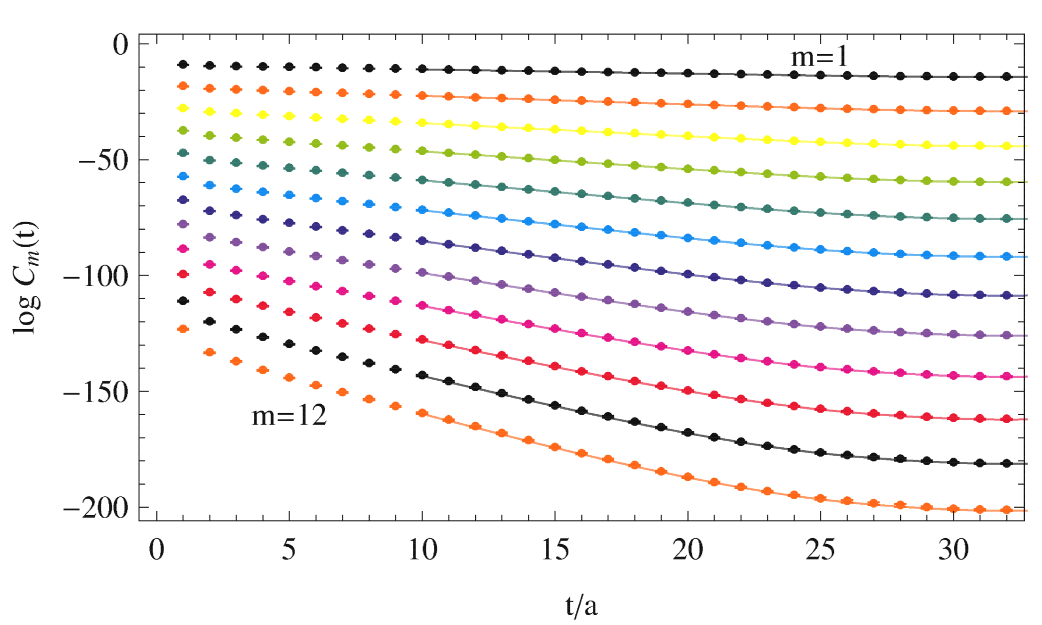}
  \caption{The logarithm of the correlators for the $m=1,\ldots,12$
    pion systems on the C1 ensemble. Also shown are the fits to these
    correlators and their uncertainties.}
  \label{fig:corr}
\end{figure}
A major issue in the present calculations is the contribution of
thermal states to two- and three- point correlations functions. Given
that the temporal extent of the lattice is fixed, one needs to go to
early Euclidean times in order to ensure that the true ground state of the
system is dominating the signal. However at early time, one must be
concerned about (forward-going) excitations that are damped at later 
Euclidean times and a careful analysis is required. By using the full
form of the expected correlator, Eq.~(\ref{eq:33}), and performing
fits to all 12 correlators, the energies $E_m$  and the multiple
overlap factors $Z_m^{(\ell)}$ can be determined. Since we
successively fit $C_1$, $C_2$, \ldots, $C_{12}$, only a single energy
is determined by each fit and to account for correlations we perform
these fits using a bootstrap procedure. The results of our fits to the 
correlators on the C1 ensemble are shown in Fig.~\ref{fig:corr}.

\begin{figure}[th]
  \centering
  \includegraphics[width=0.37\columnwidth]{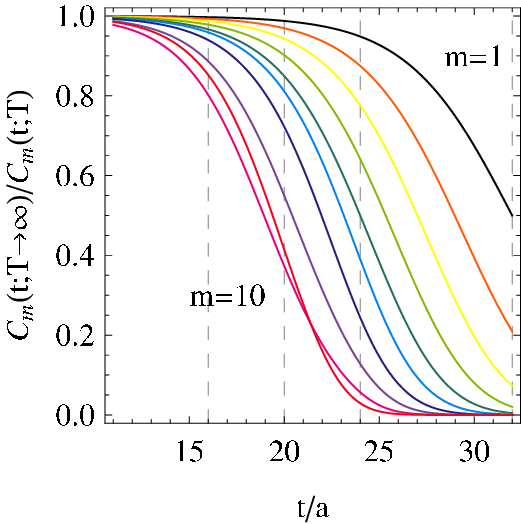}
\qquad\qquad  \includegraphics[width=0.37\columnwidth]{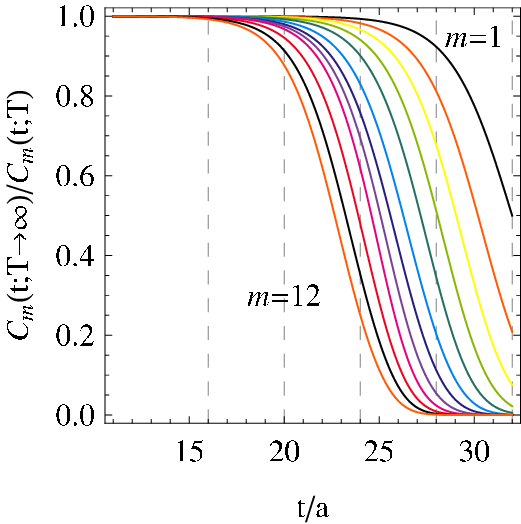}
\caption{The ratio of the zero temperature reconstruction of the
  correlators for $m=1,\ldots, 12$ pion systems to the fitted (finite
  T) correlators. Data are from the C1 (left) and C3 (right)
  ensembles. When this ratio deviates from unity, thermal effects are
  making significant contributions to the correlator.}
  \label{fig:infT}
\end{figure}

The extracted bootstrap sets of parameters can then be used to
reconstruct the correlator in the limit of infinite temporal extent by
removing all thermal contributions. In Fig.~\ref{fig:infT}, the ratios
of these zero temperature extrapolation of the correlators to the
original correlators are shown as a function of Euclidean time for the
$m=1,\ldots,12$ pion systems (again for the C1 ensemble). It is clear
from this figure that for source--sink separations beyond $t\sim20$,
it will prove difficult to extract the $m$-pion matrix element from
the corresponding three-point function as the desired contribution
will be suppressed relative to the thermal contributions. This is also
the case for the C3 ensemble (where $m_\pi T \sim 20$) for larger
numbers of pions as shown in the right panel of the figure.

\section{Quark momentum fraction}

\begin{figure}[h]
  \centering
  \includegraphics[width=0.7\columnwidth]{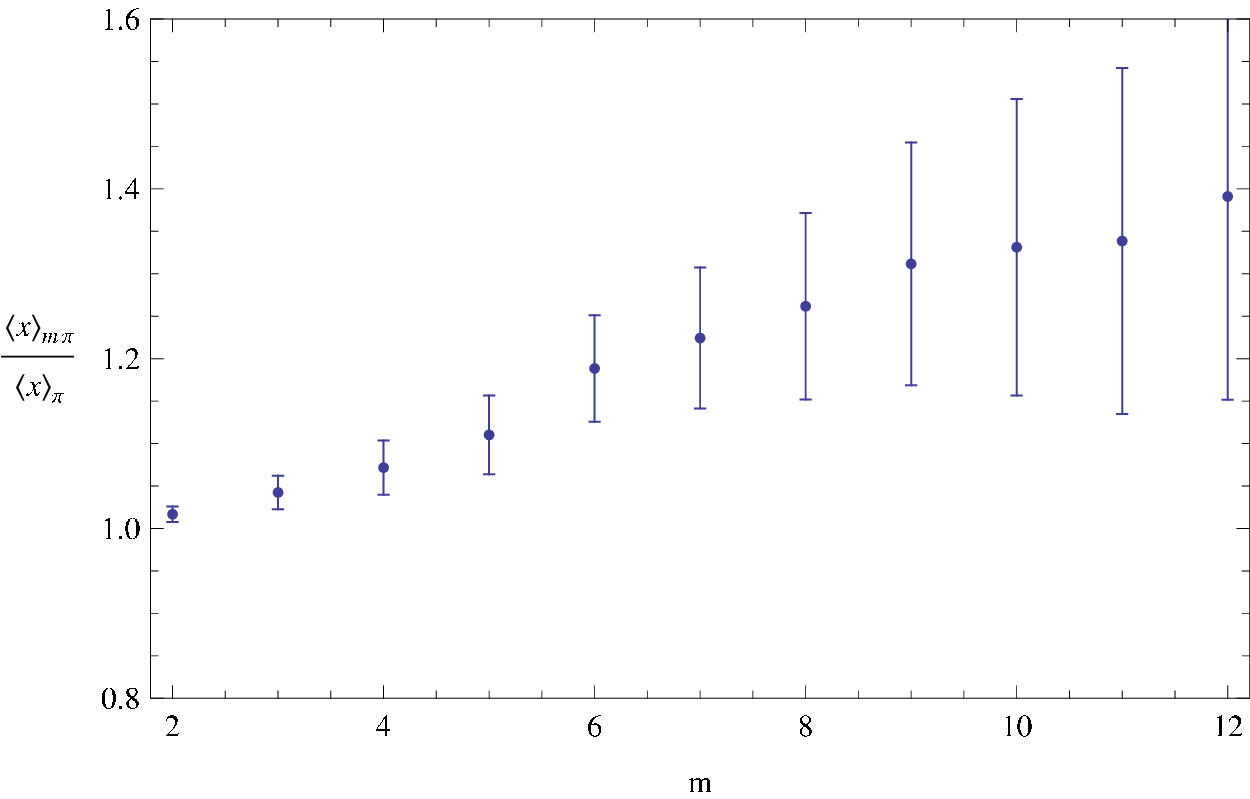}
  \caption{Extracted ratio of pion momentum fraction in an $m$-pion
    system to that in a single pion, $\langle x\rangle_{m \pi}/\langle
    x\rangle_\pi$ for the C3 ensemble as a function of the number of
    pions in the system.}
  \label{fig:x}
\end{figure}

In order to perform extractions of the matrix elements, we use the
bootstrap list of the $m$-pion energies and overlap factors obtained 
from the two-point functions and input them into the expected spectral
decomposition of the three-point function, Eq.~(\ref{eq:44}). Using
multiple different source-sink separations, we then fit the
parameters $\langle {\cal O}_m^{(n)}\rangle$ to the three point data
under the bootstrap procedure.
In Fig.~\ref{fig:x}, we show preliminary results for the the
dependence of $\langle x\rangle_{m\pi}$ on the density of the pion gas
for the C3 ensemble. Mild dependence on the density of pions is
observed but further work remains to better quantify systematic
effects and to study the dependence on quark masses and the continuum
and infinite volume limits.

\section{Discussion}
It is clear from our preliminary studies presented here that the
investigation of multi-hadron matrix elements is a challenging task in
lattice QCD. While the current investigations focus on multi-pion
systems, similar techniques in principle allow access to matrix
elements in nuclei. With the recent observation of bound light nuclei
in quenched QCD \cite{Yamazaki:2009ua,Yamazaki:2011nd} and QCD at
unphysical quark masses
\cite{Beane:2009gs,Beane:2010hg,Inoue:2010es,Beane:2011iw}, further
studies are warranted. However, the large statistics needed in lattice
studies of few-body nuclei make this a daunting task for the
future. On a more positive note, the thermal contamination that has
hampered then current investigations is particularly vexing as
$E_{m-\ell} +E_\ell <E_m$ for these systems and in light nuclei, where
the binding per nuclei increases, this may not be such a significant
issue.

\section*{Acknowledgments}
\noindent
We thank D. B. Kaplan, S. Meinel, K.~Orginos, M. J. Savage and
J.~Zanotti for useful discussions, the MILC, NPLQCD and LHP
collaborations for access to gauge configurations and propagators, and
R.~Edwards and B.~Joo for the development of the {\tt chroma} library
\cite{Edwards:2004sx}.  The work of HWL was supported by the
U.S.~Department.~of Energy (DOE) grant DE-FG03-97ER4014, and that of
WD in part by Jefferson Science Associates, LLC under DOE contract
No. DE-AC05-06OR-23177, DOE grants DE-FG02-04ER41302 and
DE-SC000-1784, and by the Jeffress Memorial Trust (J-968). Computing
support was provided by NERSC (DE-AC02-05CH11231) and the Hyak cluster at
the University of Washington eScience Institute, using hardware
awarded by NSF grant PHY-09227700.


\begin{thebibliography}{99}

\expandafter\ifx\csname natexlab\endcsname\relax\def\natexlab#1{#1}\fi
\expandafter\ifx\csname bibnamefont\endcsname\relax
  \def\bibnamefont#1{#1}\fi
\expandafter\ifx\csname bibfnamefont\endcsname\relax
  \def\bibfnamefont#1{#1}\fi
\expandafter\ifx\csname citenamefont\endcsname\relax
  \def\citenamefont#1{#1}\fi
\expandafter\ifx\csname url\endcsname\relax
  \def\url#1{\texttt{#1}}\fi
\expandafter\ifx\csname urlprefix\endcsname\relax\def\urlprefix{URL
}\fi
\providecommand{\bibinfo}[2]{#2}
\providecommand{\eprint}[2][]{\url{#2}}


\bibitem{Aubert:1983xm}
\bibinfo{author}{\bibfnamefont{J.}~\bibnamefont{Aubert}} \bibnamefont{et~al.},
  \bibinfo{journal}{Phys.Lett.} \textbf{\bibinfo{volume}{B123}},
  \bibinfo{pages}{275} (\bibinfo{year}{1983}).

\bibitem{Norton:2003cb}
\bibinfo{author}{\bibfnamefont{P.}~\bibnamefont{Norton}},
  \bibinfo{journal}{Rept.Prog.Phys.} \textbf{\bibinfo{volume}{66}},
  \bibinfo{pages}{1253} (\bibinfo{year}{2003}).

\bibitem{Detmold:2005gg}
\bibinfo{author}{\bibfnamefont{W.}~\bibnamefont{Detmold}} \bibnamefont{and}
  \bibinfo{author}{\bibfnamefont{C.}~\bibnamefont{Lin}},
  \bibinfo{journal}{Phys.Rev.} \textbf{\bibinfo{volume}{D73}},
  \bibinfo{pages}{014501} (\bibinfo{year}{2006}).

\bibitem{Detmold:2011kw}
\bibinfo{author}{\bibfnamefont{W.}~\bibnamefont{Detmold}} \bibnamefont{and}
  \bibinfo{author}{\bibfnamefont{B.}~\bibnamefont{Smigielski}},
  \bibinfo{journal}{Phys.Rev.} \textbf{\bibinfo{volume}{D84}},
  \bibinfo{pages}{014508} (\bibinfo{year}{2011}).

\bibitem{Beane:2007es}
\bibinfo{author}{\bibfnamefont{S.~R.} \bibnamefont{Beane}}
  \bibnamefont{et~al.}, \bibinfo{journal}{Phys.Rev.Lett.}
  \textbf{\bibinfo{volume}{100}}, \bibinfo{pages}{082004}
  (\bibinfo{year}{2008}).

\bibitem{Detmold:2008fn}
\bibinfo{author}{\bibfnamefont{W.}~\bibnamefont{Detmold}} \bibnamefont{et~al.},
  \bibinfo{journal}{Phys.Rev.} \textbf{\bibinfo{volume}{D78}},
  \bibinfo{pages}{014507} (\bibinfo{year}{2008}).

\bibitem{Detmold:2010au}
\bibinfo{author}{\bibfnamefont{W.}~\bibnamefont{Detmold}} \bibnamefont{and}
  \bibinfo{author}{\bibfnamefont{M.~J.} \bibnamefont{Savage}},
  \bibinfo{journal}{Phys.Rev.} \textbf{\bibinfo{volume}{D82}},
  \bibinfo{pages}{014511} (\bibinfo{year}{2010}).

\bibitem{QD}
\bibinfo{author}{\bibfnamefont{D.~H.} \bibnamefont{Bailey}},
  \bibinfo{author}{\bibfnamefont{Y.}~\bibnamefont{Hida}},
  \bibinfo{author}{\bibfnamefont{X.~S.} \bibnamefont{Li}}, \bibnamefont{and}
  \bibinfo{author}{\bibfnamefont{B.}~\bibnamefont{Thompson}},
  \bibinfo{note}{{\tt http://crd-legacy.lbl.gov/~dhbailey/mpdist/}}.


\bibitem{Bernard:2001av}
\bibinfo{author}{\bibfnamefont{C.~W.} \bibnamefont{Bernard}}
  \bibnamefont{et~al.}, \bibinfo{journal}{Phys.Rev.}
  \textbf{\bibinfo{volume}{D64}}, \bibinfo{pages}{054506}
  (\bibinfo{year}{2001}).

\bibitem{Hasenfratz:2001hp}
\bibinfo{author}{\bibfnamefont{A.}~\bibnamefont{Hasenfratz}} \bibnamefont{and}
  \bibinfo{author}{\bibfnamefont{F.}~\bibnamefont{Knechtli}},
  \bibinfo{journal}{Phys.Rev.} \textbf{\bibinfo{volume}{D64}},
  \bibinfo{pages}{034504} (\bibinfo{year}{2001}).

\bibitem{Kaplan:1992bt}
\bibinfo{author}{\bibfnamefont{D.~B.} \bibnamefont{Kaplan}},
  \bibinfo{journal}{Phys.Lett.} \textbf{\bibinfo{volume}{B288}},
  \bibinfo{pages}{342} (\bibinfo{year}{1992}).

\bibitem{Furman:1994ky}
\bibinfo{author}{\bibfnamefont{V.}~\bibnamefont{Furman}} \bibnamefont{and}
  \bibinfo{author}{\bibfnamefont{Y.}~\bibnamefont{Shamir}},
  \bibinfo{journal}{Nucl.Phys.} \textbf{\bibinfo{volume}{B439}},
  \bibinfo{pages}{54} (\bibinfo{year}{1995}).

\bibitem{Albanese:1987ds}
\bibinfo{author}{\bibfnamefont{M.}~\bibnamefont{Albanese}}
  \bibnamefont{et~al.}, \bibinfo{journal}{Phys.Lett.}
  \textbf{\bibinfo{volume}{B192}}, \bibinfo{pages}{163} (\bibinfo{year}{1987}).

\bibitem{Beane:2011zm}
\bibinfo{author}{\bibfnamefont{S.}~\bibnamefont{Beane}} \bibnamefont{et~al.}, \eprint{1108.1380}.

\bibitem{WalkerLoud:2008bp}
\bibinfo{author}{\bibfnamefont{A.}~\bibnamefont{Walker-Loud}}
  \bibnamefont{et~al.}, \bibinfo{journal}{Phys.Rev.}
  \textbf{\bibinfo{volume}{D79}}, \bibinfo{pages}{054502}
  (\bibinfo{year}{2009}).


\bibitem{Yamazaki:2009ua}
\bibinfo{author}{\bibfnamefont{T.}~\bibnamefont{Yamazaki}},
  \bibinfo{author}{\bibfnamefont{Y.}~\bibnamefont{Kuramashi}},
  \bibnamefont{and} \bibinfo{author}{\bibfnamefont{A.}~\bibnamefont{Ukawa}},
  \bibinfo{journal}{Phys.Rev.} \textbf{\bibinfo{volume}{D81}},
  \bibinfo{pages}{111504} (\bibinfo{year}{2010}).
 
\bibitem{Yamazaki:2011nd}
\bibinfo{author}{\bibfnamefont{T.}~\bibnamefont{Yamazaki}},
  \bibinfo{author}{\bibfnamefont{Y.}~\bibnamefont{Kuramashi}},
  \bibnamefont{and} \bibinfo{author}{\bibfnamefont{A.}~\bibnamefont{Ukawa}},
  \bibinfo{journal}{Phys.Rev.} \textbf{\bibinfo{volume}{D84}},
  \bibinfo{pages}{054506} (\bibinfo{year}{2011}).


\bibitem{Beane:2009gs}
\bibinfo{author}{\bibfnamefont{S.~R.} \bibnamefont{Beane}}
  \bibnamefont{et~al.}, \bibinfo{journal}{Phys.Rev.}
  \textbf{\bibinfo{volume}{D80}}, \bibinfo{pages}{074501}
  (\bibinfo{year}{2009}).

\bibitem{Beane:2010hg}
\bibinfo{author}{\bibfnamefont{S.}~\bibnamefont{Beane}} \bibnamefont{et~al.},
  \bibinfo{journal}{Phys.Rev.Lett.} \textbf{\bibinfo{volume}{106}},
  \bibinfo{pages}{162001} (\bibinfo{year}{2011}{\natexlab{b}}).

\bibitem{Inoue:2010es}
\bibinfo{author}{\bibfnamefont{T.}~\bibnamefont{Inoue}} \bibnamefont{et~al.},
  \bibinfo{journal}{Phys.Rev.Lett.} \textbf{\bibinfo{volume}{106}},
  \bibinfo{pages}{162002} (\bibinfo{year}{2011}).



\bibitem{Beane:2011iw}
\bibinfo{author}{\bibfnamefont{S.}~\bibnamefont{Beane}}
\bibnamefont{et~al.},
 \eprint{1109.2889}.

\bibitem{Edwards:2004sx}
\bibinfo{author}{\bibfnamefont{R.~G.} \bibnamefont{Edwards}} \bibnamefont{and}
  \bibinfo{author}{\bibfnamefont{B.}~\bibnamefont{Joo}},
  \bibinfo{journal}{Nucl.Phys.Proc.Suppl.} \textbf{\bibinfo{volume}{140}},
  \bibinfo{pages}{832} (\bibinfo{year}{2005}).

\end{thebibliography}
\end{document}